\documentstyle[12pt]{article}

\newcommand{\br}{{\bf r}}
\newcommand{\be}{\begin{equation}}
\newcommand{\ee}{\end{equation}}
\newcommand{\dgr}{\dagger}
\newcommand{\dlt}{\delta}
\newcommand{\prt}{\partial}
\newcommand{\vp}{\varphi}
\newcommand{\om}{\omega}
\newcommand{\bt}{\beta}
\newcommand{\al}{\alpha}
\newcommand{\Dlt}{\Delta}
\newcommand{\ra}{\rightarrow}
\newcommand{\gm}{\gamma}
\newcommand{\ep}{\varepsilon}

\newcommand{\ov}{\overline}

\begin{document}

\begin{center}
{\Large{\bf Resonant Bose Condensate: Analog of Resonant Atom} \\ [5mm]
V.I. Yukalov$^{1,2}$, E.P. Yukalova$^{2,3}$, and V.S. Bagnato$^2$} \\ [3mm]

{\it
$^1$Bogolubov Laboratory of Theoretical Physics \\
Joint Institute for Nuclear Research, Dubna 141980, Russia \\ [2mm]

$^2$Instituto de Fisica de S\~ao Carlos, Universidade de S\~ao Paulo\\
Caixa Postal 369, S\~ao Carlos, S\~ao Paulo 13560-970, Brazil \\ [2mm]

$^3$Department of Computational Physics, Laboratory of Information
Technologies \\
Joint Institute for Nuclear Research, Dubna 141980, Russia}

\end{center}

\vskip 2cm

\begin{abstract}

The resonant formation of nonlinear coherent modes in trapped 
Bose-Einstein condensates is studied. These modes represent 
nonground-state Bose condensates. The methods of describing the spectrum 
of the nonlinear modes are discussed. The latter can be created by 
modulating the trapping potential with a frequency tuned close to the 
transition frequency between the two chosen modes. The requirement that 
the transition amplitudes be smaller than the transition frequency 
implies a constraint on the number of particles that can be transferred 
to an excited mode. The resonant Bose condensate serves as a collective 
analog of a single resonant atom. Such a condensate, displaying the coherent
resonance, possesses several interesting features, among which are: mode 
locking, critical dynamics, interference patterns, interference current, 
atomic squeezing, and multiparticle entanglement. 

\end{abstract}

\newpage

\section{Introduction}

Dilute atomic gases, trapped and cooled down to temperatures when almost 
all atoms are in the Bose-condensed state, are described by the 
Gross-Pitaevskii equation (see reviews [1--3]). The mathematical structure 
of the latter is that of the nonlinear Schr\"odinger equation which, due 
to the presence of the trapping potential, possesses a discrete spectrum. 
The equilibrium Bose-Einstein condensate corresponds to the ground state
associated with the lowest energy level of the spectrum. If not the ground 
but an excited state would be macroscopically populated, this would 
correspond to a {\it nonground-state} Bose condensate. The possibility of 
creating such a nonequilibrium condensate was advanced in Ref. [4]. This 
can be done by applying an alternating field with a frequency tuned to 
the transition frequency between the ground state and a chosen excited 
state. The latter states are called the nonlinear coherent modes and they 
are described by the stationary solutions to the Gross-Pitaevskii 
equation. The properties of these modes have been considered 
theoretically in several publications [4--13] and a nonlinear dipole mode 
was observed experimentally [14]. A known example of such a mode is a 
vortex that can be formed by means of a rotating laser spoon [15,16].

The aim of the present publication is threefold. First, we give a survey 
of the problems in characterizing the nonlinear coherent modes, 
especially in the accurate calculations of their spectrum. Second, we 
stress the analogy of the resonant Bose condensate with a resonant atom. 
And, third, we show that, because of its coherent collective nature, the 
resonant Bose condensate possesses a number of properties distinguishing 
it from a single resonant atom. We describe several interesting novel 
effects that can be observed in a nonequilibrium Bose condensate.

\section{Nonlinear Coherent Modes}

First of all, it is necessary to show how one could get an accurate 
description of nonlinear coherent modes. We assume that the system of $N$ 
Bose atoms is confined in a trap, with a trapping potential $U(\br)$ 
tending to infinity at large $r\equiv|\br|$,
$$
U(\br) \rightarrow \infty \qquad (r\ra \infty)\; .
$$
Due to the presence of this confining potential, the spectrum of the 
stationary Gross-Pitaevskii equation is discrete, being defined by the 
eigenvalue problem
\be
\label{1}
\hat H[\vp_n]\; \vp_n(\br) = E_n\; \vp_n(\br) \; ,
\ee
with the nonlinear Hamiltonian
\be
\label{2}
\hat H[\vp] \equiv \; -\; \frac{\hbar^2{\nabla}^2}{2m_0} +
U(\br) + N \int \Phi(\br-\br')\; |\vp(\br')|^2\; d\br' \; ,
\ee
where $m_0$ is atomic mass and $\Phi(\br)$ is a potential of interatomic 
interactions. The eigenfunction $\vp_n(\br)$, labelled by a multi-index 
$n$, is a {\it nonlinear coherent mode}. It is assumed to be normalized 
to unity, $||\vp_n||=1$. Atomic interactions for dilute trapped gases are 
described by the Fermi contact potential
\be
\label{3}
\Phi(\br) = A_s\; \dlt(\br) \; , \qquad A_s \equiv 4\pi\hbar^2\;
\frac{a_s}{m_0} \; ,
\ee
with $a_s$ being the $s$-wave scattering length. The trapping potential 
is commonly presented by the harmonic oscillator form
\be
\label{4}
U(\br) = \frac{m_0}{2}\left ( \om_x^2\; r_x^2 + \om_y^2\; r_y^2 +
\om_z\; r_z^2 \right ) \; .
\ee
Here, we shall consider the single-well potential (4), though the 
double-well and more complicated potentials [6,11,13] are possible, 
including periodic potentials corresponding to optical lattices [17--31].

In the case of cylindric symmetry, when $\om_x=\om_y\equiv\om_r$, it is 
convenient to introduce the notation
\be
\label{5}
\nu \equiv \frac{\om_z}{\om_r} \; , \qquad
l_r \equiv \sqrt{\frac{\hbar}{m_0\om_r}}
\ee
for the frequency ratio $\nu$ and oscillator length $l_r$, which allows 
the usage of the dimensionless spatial variables
$$
r_\perp \equiv \frac{\sqrt{r_x^2+r_y^2}}{l_r}\; , \qquad
z\equiv \frac{r_z}{l_r} \; .
$$
And the quantity
\be
\label{6}
g \equiv 4\pi\; \frac{a_s}{l_r}\; N
\ee
is a dimensionless coupling parameter. Also, let us define the dimensionless 
Hamiltonian and the wave function, respectively,
$$
\hat H \equiv \frac{\hat H[\vp]}{\hbar \om_r} \; , \qquad 
\psi(r_\perp,\vp,z) \equiv l_r^{3/2}\; \vp(\br) \; .
$$
Then Eq. (2) reduces to
\be
\label{7}
\hat H = -\; \frac{1}{2}\; {\nabla}^2 + \frac{1}{2} \left (
r_\perp^2 + \nu^2 z^2 \right ) + g|\psi|^2 \; ,
\ee
where
$$
{\nabla}^2 = \frac{\prt^2}{\prt r_\perp^2} + \frac{1}{r_\perp}\;
\frac{\prt}{\prt r_\perp} + \frac{1}{r_\perp^2}\;
\frac{\prt^2}{\prt\vp^2} + \frac{\prt^2}{\prt z^2} \; .
$$
The eigenproblem (1) takes the form $\hat H\psi_{nmj} = E_{nmj}\psi_{nmj}$,
in which $n=0,1,2,\ldots$ is the radial quantum number, $m=0,\pm 1,\pm 
2,\ldots$ is the azimuthal quantum number, and $j=0,1,2,\ldots$ is the 
axial quantum number.

The equilibrium Bose-Einstein condensate corresponds to the ground-state 
solution to the stationary Gross-Pitaevskii equation, when $n=m=j=0$. This
equation is nonlinear and does not allow for an exact solution. If the 
coupling parameter (6) were small, $g\ra 0$, perturbation theory would be 
admissible. And in the opposite limit of asymptotically strong coupling, 
$g\ra\infty$, the Thomas-Fermi approximation could be used (see [1--3]). 
Here we demonstrate how one can construct an accurate solution for the 
ground state, which is valid for arbitrary coupling parameters 
$g\in[0,\infty)$. For simplicity, we consider the spherical trap, when 
$\nu=1$. Then the ground state $\psi_0(r)\equiv \psi_{000}(r)$ depends on 
the spherical variable $r\equiv\sqrt{r_\perp^2 + z^2}$.

To find an accurate approximate solution of a nonlinear differential 
equation, we employ the {\it self-similar approximation theory} [32--36], 
in the variant [36--38] designed for constructing {\it self-similar 
crossover approximants} satisfying the corresponding asymptotic 
conditions. To this end, we notice that at short radius $r\ra 0$, when 
the nonlinear term in the Hamiltonian dominates, one has
\be
\label{8}
\psi_0(r) \simeq c_0 + c_2 r^2 + c_4 r^4 \qquad (r\ra 0)\; .
\ee
And at large $r\ra\infty$, where the harmonic term in Eq. (7) becomes 
dominant, the wave function tends to the Gaussian form
\be
\label{9}
\psi_0(r) \simeq C\exp\left ( -\; \frac{1}{2}\; r^2 \right ) \qquad
(r\ra\infty) \; .
\ee
The self-similar crossover approximant, interpolating between the 
asymptotic limits (8) and (9) is
\be
\label{10}
\psi_0^*(r) = C\exp\left ( - \; \frac{1}{2}\; r^2 \right ) \;
\exp\left\{ ar^2 \exp(-br^2)\right \} \; .
\ee
The coefficients $a$ and $b$ are obtained by expanding function (10) in
powers of $r$ and substituting this expansion in the stationary 
Gross-Pitaevskii equation, which yields
$$
a=\frac{1}{2} + \frac{1}{3} \left ( gC^2 - E_0\right ) \; , \qquad
b=\frac{1}{10a} \; ( 1-2a)\left ( E_0 - 1 + 2a\right ) -\;
\frac{1}{20a} \; .
$$
The ground-state energy $E_0$ and the normalization constant $C$ are 
defined by the consistency conditions
\be
\label{11}
4\pi \int_0^\infty \psi_0(r) \; \hat H\psi_0(r)\; r^2\; dr = E_0 \; , \qquad
4\pi \int_0^\infty \psi_0^2(r) r^2 \; dr = 1 \; .
\ee
The accuracy of an approximate solution can be characterized by 
calculating the {\it local residual}
\be
\label{12}
R(r) \equiv \sqrt{4\pi}\; r \left ( \hat H - E_0 \right ) \psi_0 (r)\; ,
\ee
showing the deviation of the radial wave function from the exact solution
at each point $r$. The behaviour of this residual for the self-similar 
approximant (10) is presented in Fig. 1 for different values of the 
coupling parameter (6), where it is compared with the residuals for the 
optimized Gaussian approximation and Thomas-Fermi approximation [3]. As 
is seen, the accuracy of the self-similar approximant (10) is essentially 
better. Figure 2 shows the atomic density
\be
\label{13}
n(r) \equiv 4\pi \psi_0^2(r)
\ee
for the self-similar approximant (10), as compared with that for the 
optimized Gaussian and Thomas-Fermi approximations.

The way of constructing an accurate approximation for the wave function 
is suitable for describing the ground state in a spherical trap, but it 
becomes rather cumbersome for nonspherical traps and for nonground states. 
Moreover, the exact shape of wave functions is of less importance than 
the related energy values. Hence it is necessary to concentrate attention 
on the calculation of the spectrum of higher nonlinear modes. For this 
purpose, the most suitable is the {\it optimized perturbation theory} 
[39--41]. This approach has been successfully employed for a variety of 
physical problems. A brief survey of the method and many citations can be 
found in Refs. [35,36]. Recently, this approach was used for calculating 
the critical temperature of Bose-Einstein condensation in a dilute Bose 
gas [42--47]. The main idea of the optimized perturbation theory is the
introduction of control functions that optimize the convergence of a 
calculational procedure [39--41]. Control functions can be introduced in 
different ways, and different calculational algorithms may be employed. 
Any kind of perturbation theory or iterative procedure can be used. The 
Rayleigh-Schr\"odinger and Dalgarno-Lewis perturbation theories can be 
invoked for the Schr\"odinger equation as well as for Green functions [48].
Semiclassical expansions are possible [49]. One may also prefer to work 
not with the Schr\"odinger equation itself but with an inverse 
Schr\"odinger equation [50].

We apply the optimized perturbation theory, with the Rayleigh-Schr\"odinger 
algorithm, to the Gross-Pitaevskii equation (1). For the spectrum, in the 
first approximation, we find
\be
\label{14}
e(g) \equiv E(g,u(g),v(g)) \; ,
\ee
with the expression
\be
\label{15}
E(g,u,v) = \frac{p}{2}\left ( u + \frac{1}{u}\right ) + \frac{q}{4}
\left ( v + \frac{\nu^2}{v} \right ) + u\; \sqrt{v}\; g I_{nmj} \; ,
\ee
in which
$$
p\equiv 2n + |m| + 1 \; , \qquad q \equiv 2j + 1 
$$
are the combinations of quantum numbers, and the integral
$$
I_{nmj} \equiv \frac{1}{u\sqrt{v}} \; \int_0^\infty r_\perp\; dr_\perp \;
\int_0^{2\pi} d\vp \;  \int_{-\infty}^{+\infty} dz \; 
|\psi_{nmj}(r_\perp,\vp,z)|^4
$$
contains the wave functions
$$
\psi_{nmj}(r_\perp,\vp,z)  =\left [ 
\frac{2n!\; u^{|m|+1}}{(n+|m|)!} \right ]^{1/2} \; r_\perp^{|m|}\;
\exp\left ( -\; \frac{u}{2}\; r_\perp^2\right ) \times
$$
$$
\times L_n^{|m|}\left ( u r_\perp^2\right ) \; 
\frac{e^{im\vp}}{\sqrt{2\pi}}\; \left ( \frac{v}{\pi}\right )^{1/4} \;
\frac{1}{\sqrt{2^j\; j!}}\; \exp\left ( -\; \frac{v}{2}\; z^2\right )\;
H_j\left ( \sqrt{v}\; z\right ) \; ,
$$
where $L_n^m(\cdot)$ is a Laguerre polynomial and $H_j(\cdot)$ is a 
Hermit polynomial. The control functions $u=u(g)$ and $v=v(g)$ are
obtained from the optimization condition
$$
\left ( \dlt u\; \frac{\prt}{\prt u} + \dlt v \; \frac{\prt}{\prt v}
\right ) \; E(g,u,v) = 0 \; ,
$$
which results in the equations
\be
\label{16}
p\left ( 1 -\; \frac{1}{u^2}\right ) + \frac{G}{p\nu}\;
\sqrt{\frac{v}{q}} = 0 \; , \qquad q \left ( 1 - \; 
\frac{\nu^2}{v^2}\right ) + \frac{uG}{p\nu\; \sqrt{vq}} =0 \; ,
\ee
in which
\be
\label{17}
G\equiv 2p\; \sqrt{q}\; I_{nmj}\; g\nu \; .
\ee
Equations (15) to (17) define the spectrum (14) for all quantum numbers
$n,\; m$, and $j$ and for arbitrary values of the coupling parameter $g$.

Instead of solving numerically Eqs. (14) to (17), we may derive explicit
analytical expressions for the spectrum of nonlinear modes by resorting 
to the technique of self-similar root approximants [36--38]. To this end,
we find the weak-coupling expansion of spectrum (14) in the form
\be
\label{18}
e(g) \simeq \sum_{n=0}^k a_n\; G^n \qquad (G\ra 0) \; ,
\ee
with the coefficients
$$
a_0 = p +\frac{Q}{2} \; , \quad a_1 = \frac{1}{2pQ^{1/2}}\; , 
\quad a_2 = -\; \frac{p+2Q}{16p^3Q^2} \; , \quad 
a_3=\frac{(p+2Q)^2}{64p^5Q^{7/2}} \qquad (Q\equiv q\nu) \; .
$$
Also, we obtain the strong-coupling expansion 
\be
\label{19}
e(g) \simeq \sum_{n=0}^k b_n\; G^{\bt_n} \qquad (G\ra\infty) \; ,
\ee
where the coefficients are given by the equalities
$$
4b_0=5\; , \qquad 4b_1=2p^2 + Q^2 \; ,
$$
$$ 
20b_2 =-3p^4 + 2p^2 Q^2 - 2Q^4 \; , \qquad
29 b_3 = 2p^6 - p^4 Q^2 - 2p^2 Q^4 + 2Q^6 \; ,
$$
$$ 
500 b_4 = -44p^8 + 22p^6 Q^2 +2p^4 Q^4 + 78 p^2 Q^6- 69 Q^8\; ,
$$
$$
12500 b_5 = 1122p^{10} - 595 p^8 Q^2 - 70p^6 Q^4 + 440 p^4Q^6 -
3640 p^2 Q^8 + 2821 Q^{10} \; ,
$$
and the powers are
$$
\bt_0=\frac{2}{5} \; \qquad \bt_1=-\;\frac{2}{5}\; , \qquad
\bt_2=-\; \frac{6}{5}\; , 
$$
$$
\bt_3=-2 \; \qquad \bt_4=-\;\frac{14}{5}\; , \qquad
\bt_5=-\; \frac{18}{5}\; .
$$
Interpolating the asymptotic limits (18) and (19) by means of the
self-similar root approximants [36--38], we obtain
\be
\label{20}
e_k^*(g) = a_0 \left (\left ( \ldots \left ( 1 + A_{k1} G
\right )^{n_{k1}} + A_{k2} G^2\right )^{n_{k_2}} + \ldots 
A_{kk} G^k\right )^{n_{kk}} \; .
\ee
Here, depending on the approximation order $k=1,2,\ldots$, we have in 
the first order
$$
A_{11}= \frac{1.746928}{a_0^{5/2}} \; , \qquad n_{11} =\frac{2}{5} \; ,
$$
in the second order
$$
A_{21}=2.533913\; \frac{( 2p^2 +Q^2)^{5/6}}{a_0^{25/6}} \; ,
\qquad A_{22}=\frac{3.051758}{a_0^5} \; , \qquad n_{21}=\frac{6}{5}\; ,
\qquad n_{22} =\frac{1}{5} \; ,
$$
and in the third order
$$
A_{31}=1.405455\; 
\frac{(8p^4+12p^2Q^2+Q^4)^{5/6}}{a_0^{125/22}(2p^2+Q^2)^{5/66}} \; ,
\qquad A_{32} =6.619620 \; \frac{(2p^2+Q^2)^{10/11}}{a_0^{75/11}} \; ,
$$
$$
A_{33} =\frac{5.331202}{a_0^{15/2}} \; , \qquad
n_{31} =\frac{6}{5} \; , \qquad n_{32} =\frac{11}{10} \; , \qquad 
n_{33} =\frac{2}{15} \; .
$$
The accuracy of approximants (20), for vortex nonlinear modes with
$n=j=0$ and different $m$, as compared to the numerical values of Eq.
(14), are illustrated in Fig. 3. As we see, the approximants $e_2^*(g)$
and $e_3^*(g)$ are quite accurate.

The analytical expression (20) makes it possible a direct analysis of 
the spectrum with respect to varying quantum numbers and the coupling
parameter. It also allows us to get a fast estimate of the critical
parameter $g_c$, and, hence, of the critical number of atoms above which
the system of atoms with attractive interactions becomes unstable.
For this purpose, we need to find the negative value of $G_c$ at which
the spectrum (20) becomes complex. To illustrate the idea, we limit
ourselves by the first approximation $e_1^*(g)$, which is complex for
$G<G_c$, with
$$
G_c=-0.572433\; a_0^{5/2} \; .
$$
This, in view of notation (17), gives
$$
g_c \approx -0.05\; 
\frac{(2p+q\nu)^{5/2}}{p\sqrt{q}\; I_{nmj}\nu} \; ,
$$
which, because of Eq. (6), yields the estimate for the critical number
of atoms
$$
N_c \approx \frac{(2p+q\nu)^{5/2}}{300p\sqrt{q}\; I_{nmj}\nu} \;
\left | \frac{l_r}{a_s}\right | \; .
$$
This estimate is of order of that obtained by means of numerical 
solution [3,4]. A more accurate value of $N_c$ can be derived from 
the consideration of the higher-order approximants (20).

\section{Resonant Bose Condensate}

Suppose that a trapped atomic gas is cooled down to temperatures when 
almost all atoms are in the Bose-Einstein condensate corresponding 
to the ground state energy $E_0$. As has been shown [4], applying an 
alternating external field, with the frequency close to the transition
frequency between the ground state and an excited mode, it is possible 
to transfer atoms from the ground to this excited mode. Thus, a nonground
state condensate can be created. The latter, by applying one more resonant 
field can be transferred to another higher mode. In this way, we may consider 
resonant transitions  between two energy levels, say $E_1$ and $E_2$, such 
that $E_1<E_2$, with the transition frequency 
\be
\label{21}
\om_{21} \equiv \frac{1}{\hbar}\; \left ( E_2 - E_1 \right ) \; .
\ee
The transition is induced by an alternating field
\be
\label{22}
\hat V =  V_1(\br)\cos\om t + V_2(\br)\sin\om t \; ,
\ee
whose frequency is close to frequency (21), so that the resonance condition
\be
\label{23}
\left | \frac{\Dlt\om}{\om} \right | \ll 1 \; , \qquad
\Dlt\om \equiv \om -\om_{21}
\ee
holds. This situation is similar to the resonant interaction of 
electromagnetic field with a single atom, though a Bose condensate is 
a many-particle system and the transition is between collective levels.

In the presence of the alternating field (22), we have to deal with the
time-dependent Gross-Pitaevskii equation
\be
\label{24}
i\hbar\; \frac{\prt}{\prt t}\; \vp(\br,t) = \left ( \hat H[\vp] +
\hat V\right ) \vp(\br,t) \; ,
\ee
with the nonlinear Hamiltonian (2). One may look for the solution to this
equation in the form of the expansion
\be
\label{25}
\vp(\br,t) = \sum_n c_n(t)\; \vp_n(\br) \;
\exp\left ( -\; \frac{i}{\hbar}\; E_n t \right )
\ee
over the nonlinear coherent modes $\vp_n(\br)$ defined by the eigenvalue 
problem (1). Note that, in general, one could look for a solution 
expanding the latter over an arbitrary basis [51]. However, the
set $\{\vp_n(\br)\}$ of the nonlinear coherent modes stands out as the sole 
natural basis, being physically distinguished from all other formally 
admissible expansion sets.

There are two types of transition amplitudes in the problem,
$$
\al_{mn} \equiv A_s\; \frac{N}{\hbar} \int |\vp_m(\br)|^2\left [
2|\vp_n(\br)|^2 - |\vp_m(\br)|^2 \right ] \; d\br \; , 
$$
\be
\label{26}
\bt_{mn} \equiv \frac{1}{\hbar} \int \vp_m^*(\br) \left [ V_1(\br) -
iV_2(\br) \right ] \; \vp_n(\br)\; d\br \; ,
\ee
the former being due to atomic interactions (3) and the latter, to the
action of the modulating field (22). In analogy with the resonance in an
atom, we need here that the transition amplitudes (26) be smaller than 
the transition frequency (21),
\be
\label{27}
\left | \frac{\al_{12}}{\om_{21}} \right | \ll 1 \; , \qquad 
\left | \frac{\al_{21}}{\om_{21}} \right | \ll 1 \; , \qquad 
\left | \frac{\bt_{12}}{\om_{21}} \right | \ll 1 \; , 
\ee
in order that the transition frequency, and the resonance as such, be well
defined. These conditions lead to the restriction
\be
\label{28}
|g\nu| < \frac{(2p^2+Q^2)^{5/4}}{14p\sqrt{q}\; I_{nmj}}
\ee
on the system parameters, which can be reformulated as the limitation
\be
\label{29}
N < \frac{(2p^2+Q^2)^{5/4}}{56\pi\nu p\sqrt{q}\; I_{nmj}}\; \left |
\frac{l_r}{a_s}\right |
\ee
on the number of particles that can be resonantly transferred to a chosen mode.
Note that the limiting value in the right-hand side of Eq. (29) is close to
$N_c$.

Under conditions (27), the Gross-Pitaevskii equation (24) reduces to the
system of equations
\be
\label{30}
i \; \frac{dc_1}{dt} = \al_{12}|c_2|^2 c_1 + 
\frac{1}{2}\;\bt_{12} c_2 e^{i\Dlt\om t} \; , \qquad
i \; \frac{dc_2}{dt} = \al_{21}|c_1|^2 c_2 + 
\frac{1}{2}\;\bt_{12}^* c_1 e^{-i\Dlt\om t} 
\ee
for the fractional population amplitudes. This reduction to an effective
two-mode system has become possible due to the resonance condition (23) and
inequalities (27). Such a situation is analogous to a resonant atom whose 
description also reduces to a two-level system. One more requirement is that
the energy levels would not be equidistant [4]. This is always valid for
electronic levels in an atom. Fortunately, owing to the nonlinear term in 
the Gross-Pitaevskii equation, this is also true for the spectrum of the
nonlinear modes. There is as well another possibility of shifting the required 
levels by a slight modification of the trapping potential [52].

Equations (30), constituting a system of four equations for real functions,
allow a further simplification. For this purpose, we may write
\be
\label{31}
c_1 = \sqrt{\frac{1-s}{2}}\; \exp( i\pi_1 t ) \; , \qquad
c_2 =\sqrt{\frac{1+s}{2}}\; \exp(i\pi_2 t) \; ,
\ee
where $\pi_1=\pi_1(t)$ and $\pi_2=\pi_2(t)$ are the phases and
\be
\label{32}
s\equiv |c_2|^2 - |c_1|^2
\ee
is the population difference. We also introduce the phase difference
\be
\label{33}
x \equiv \pi_2 - \pi_1 +\gm + \Dlt \om t \; ,
\ee
employ the notation
$$
\al_0 \equiv \frac{1}{2}\; (\al_{12}+\al_{21} ) \; , \qquad \bt_{12} \equiv
\bt e^{i\gm} \; , \qquad \bt \equiv |\bt_{12}| \; , \qquad \dlt \equiv
\Dlt\om + \frac{1}{2}\; (\al_{12} - \al_{21} ) \; ,
$$
and define the dimensionless parameters
\be
\label{34}
b\equiv \frac{\bt}{\al_0} \; , \qquad \ep \equiv \frac{\dlt}{\al_0} \; .
\ee
Then Eqs. (30) can be reorganized to the system of two equations
\be
\label{35}
\frac{ds}{dt} = -\bt\sqrt{1-s^2} \; \sin x \; , \qquad
\frac{dx}{dt} = \al_0 s + \frac{\bt s}{\sqrt{1-s^2}}\; \cos x +\dlt
\ee
for the real functions (32) and (33), with the initial conditions $s_0=s(0)$ 
and $x_0=x(0)$. Here we have considered the resonant field of the form (22),
with the amplitudes depending on the spatial variable $\br$ but not depending
on time $t$. In general, we could keep in mind that the amplitudes $V_i(\br,t)$
are slow functions of time, that is, varying much slower that the oscillation
period $2\pi/\om$. If so, the value $\bt$ in Eqs. (35) would be a function of 
time, which would permit us to study the behaviour of $s(t)$ and $x(t)$ under
the influence of resonant pulses. This would enrich the variety of admissible
solutions yielding the appearance of different transient phenomena, analogously
to the case of transient coherent phenomena in systems of resonant atoms [53].

\section{Dynamic Resonant Effects}

The resonant Bose condensate, though being analogous to a resonant atom,
possesses, because of its collective nonlinear nature, several features that
make it rather different from the latter. Some of the interesting effects
exhibited by the resonant Bose condensate, are surveyed below.

\subsection{Mode locking}

We shall say that the modes are locked if the fractional populations 
$|c_i(t)|^2$ oscillate in time so that they do not cross the line
$|c_1|^2=|c_2|^2=1/2$. Depending on initial conditions, it may be that
$|c_1|^2>1/2$ and $|c_2|^2<1/2$, or that $|c_2|^2>1/2$ while $|c_1|^2<1/2$.
With regard to the population difference (32), the mode locking means that,
depending on initial conditions, either
\be
\label{36}
-1\leq s(t) \leq 0
\ee
or
\be
\label{37}
0\leq s(t) \leq 1 \; .
\ee
Such a mode-locked regime exists for some region of the parameters (34), in
particular, when both $b$ and $\ep$ are small [4].

\subsection{Critical dynamics}

When the amplitude of the resonant field $b$ or detuning $\ep$ increase in
the magnitude, they may reach the values at which the dynamics of the
fractional populations experience dramatic changes [5,8,12]. This happens
on the critical line of the parametric manifold. More precisely, for each 
given set of initial conditions, $s_0$, $x_0$, there is a separate critical
line on the plane $\{ b,\ep\}$, described by the equation
\be
\label{38}
\frac{s_0^2}{2} \; - b \sqrt{1-s_0^2}\; \cos x_0 + \ep s_0 = |b| \; .
\ee
The critical behaviour of fractional populations, when crossing the critical
line (38), has been investigated numerically [5,8,12]. The dynamic critical
phenomena appear when the motion changes from the mode-locked type to 
mode-unlocked type. This happens when a given starting point of a trajectory 
is crossed by a saddle separatrix on the plane of the variables $s(t)$, 
$x(t)$. Thus, for the starting point $s_0=-1$, $x_0=0$, and the detuning
$\ep=-0.1$, the critical pumping amplitude, defined by Eq. (38), is 
$b_c=0.39821$. The motion, starting at $s_0=-1$, $x_0=0$, is mode locked
for $b<b_c$, and becomes mode unlocked for $b>b_c$. The phase portrait
for $b=0.51>b_c$ is shown in Fig. 4, where it is seen that the motion 
beginning at $s_0=-1$, $x_0=0$ is mode unlocked, i.e. the trajectory lies
in the interval $-1\leq s\leq 1$.

Another way of explaining the change from the mode-locked regime to the 
mode-unlocked one can be as follows. Let us introduce the function
\be
\label{39}
h \equiv 2c_1^* c_2 e^{i(\Dlt\om t + \gm)} = \sqrt{1-s^2}\; e^{ix} \; .
\ee
And let us measure time in units of $\al_0$. Then Eqs. (30) can be presented 
in the form
\be
\label{40}
\frac{dh}{dt} = i(s+\ep)h + ibs \; , \qquad
\frac{ds}{dt} = \frac{i}{2}\; b \left ( h-h^*\right ) \; .
\ee
From the first of these equations, we have
$$
h = -\; \frac{1}{s+\ep} \left ( bs + i\; \frac{dh}{dt}\right ) \; .
$$
Substituting this in the second of Eqs. (40) gives
$$
2(s+\ep)\; \frac{ds}{dt} = b\; \frac{d}{dt}\left ( h + h^*\right ) \; .
$$
The latter equation is easily integrated yielding
$$
(s+\ep)^2 - b\left ( h+h^*\right )= k_0 \; ,
$$
with the integration constant
$$
k_0\equiv (s_0+\ep)^2 - b \left ( h_0 + h_0^*\right ) = (s_0+\ep)^2 -
2b\; \sqrt{1-s_0^2}\; \cos x_0 \; .
$$
Differentiating the second of Eqs. (40), we get
$$
\frac{d^2s}{dt^2} = - \; \frac{b}{2}\; (s+\ep)\left ( h +h^*\right )
-bs^2 \; ,
$$
which can be transformed to 
$$
\frac{d^2s}{dt^2} = -\; \frac{1}{2}\; (s+\ep)^3 + \frac{k_0}{2}\; (s+\ep) - b^2s \; .
$$
From here, for the shifted population difference
$$
\ov s \equiv s+\ep \; ,
$$
we find
\be
\label{41}
\left ( \frac{d\ov s}{dt}\right )^2  = -\; \frac{1}{4} \; \ov s^4 + 
\frac{k_1}{2}\; \ov s^2 + 2b^2 \ep\ov s + k_2 \; ,
\ee
with the integration constants
$$
k_1\equiv k_0 - 2b^2 \; , \qquad
k_2 \equiv \dot{s}_0^2 + \frac{1}{4}\; \ov s_0^4 -\; \frac{k_1}{2}\;
\ov s_0^2 - 2b^2 \ep\ov s_0 \; ,
$$
where
$$
\dot{s}_0 = \frac{i}{2}\; b \left ( h_0 - h_0^* \right ) = - b\;
\sqrt{1-s_0^2} \; \sin x_0 \; .
$$
The exact solution to Eq. (41) can be expressed [54] as a ratio of the Jakobi
elliptic functions [55].

To simplify the consideration, let us set the zero detuning $\ep=0$. Then 
$\ov s = s$. For the integration constants, we get
$$
k_1 = s_0^2 - 2b\; \sqrt{1-s_0^2}\; \cos x_0  - 2b^2 \; , \qquad
k_2 = \frac{1}{4}\; s_0^4 -\; \frac{k_1}{2}\; s_0^2 + b^2 \left ( 1-s_0^2
\right ) \sin^2 x_0 \; .
$$
Equation (41) reduces to
\be
\label{42}
\left ( \frac{ds}{dt} \right )^2 = -\; \frac{1}{4}\; s^4 +
\frac{k_1}{2}\; s^2 + k_2 \; .
\ee
The polynomial in the right-hand side of Eq. (42) has the roots
\be
\label{43}
s^2 = k_1 \pm \sqrt{k_1^2 + 4k_2} \; ,
\ee
which prescribe the admissible region of varying $s$. For instance, in the case
of the initial conditions $s_0=\pm 1$, we obtain the roots
\be
\label{44}
s_{1,2} = \pm 1 \; , \qquad s_{3,4} =\pm \sqrt{1-4b^2}\; .
\ee
The left-hand side of Eq. (42) is always nonnegative, hence it should be that
\be
\label{45}
1-4b^2 \leq s^2 \leq 1 \; .
\ee
The analysis of the inequalities (45) tells as that for $b^2\leq 1/4$ the 
mode-locking regime is realized, since the motion is locked in one of the 
regions
\begin{eqnarray}
\left. \begin{array}{c}
-1 \leq s \leq -(1-4b^2) \\
\\
1-4b^2 \leq s \leq 1 \end{array} \right \} \left ( b^2 < 
\frac{1}{4}\right ) 
\end{eqnarray}
depending on initial conditions $s_0=\pm 1$. But for higher $b^2$, the 
motion becomes unlocked, with the trajectory wandering in the whole admissible
region
\be
\label{47}
-1 \leq s\leq 1 \qquad \left ( b^2 > \frac{1}{4}\right ) \; .
\ee
The change of the mode-locked regime to the mode-unlocked one happens at the
critical value $b_c^2 = 1/4$.

\subsection{Interference effects}

Since the spatial distribution of two different nonlinear modes $\vp_1(\br)$ and
$\vp_2(\br)$ is different, there appears the {\it interference pattern} described 
by
\be
\label{48}
\rho_{int}(\br,t) \equiv \rho(\br,t) -\rho_1(\br,t) -\rho_2(\br,t) \; ,
\ee
where
$$
\rho(\br,t) = \left | c_1(t)\vp_1(\br) e^{-iE_1t/\hbar} + c_2(t)\vp_2(\br)
e^{-iE_2t/\hbar} \right |^2 \; ,
$$
$$ 
\rho_j(\br,t)= |c_j(t)\vp_j(\br)|^2 \qquad (j=1,2) \; .
$$
The properties of the pattern (48) were studied in Ref. [12].

Different spatial shapes of the nonlinear modes and their different related 
energies result in the appearance of the {\it interference current}
\be
\label{49}
{\bf j}_{int}(\br,t) \equiv {\bf j}(\br,t) - {\bf j}_1(\br,t) - 
{\bf j}_2(\br,t) \; ,
\ee
in which ${\bf j}(\br,t)$ is the total current in the system, while 
${\bf j}_i(\br,t)$, with $i=1,2$ are the partial currents of the corresponding
modes [12].

\subsection{Atomic squeezing}

The evolution equations (30) for the effective two-mode system may be rewritten in 
the spin representation as the equations for the collective spin operators
$$
S_\al \equiv \sum_{i=1}^N S_i^\al\; , \qquad S_\pm \equiv S_x \pm iS_y \; ,
$$
where $\al=x,y,z$. Thence, we may consider atomic squeezing connected with the spin
operators, because of which it is called spin squeezing [56--58]. Thus, the squeezing
of $S_z$ with respect to $S_\pm$ is described by the {\it squeezing factor}
\be
\label{50}
Q_z \equiv \frac{2\Dlt^2(S_z)}{|<S_\pm>|} \; ,
\ee
in which $\Dlt^2(S_z)\equiv<S_z^2>-<S_z>^2$. For the resonant Bose condensate,
we find
\be
\label{51}
Q_z =\sqrt{1-s^2} \; ,
\ee
with $s$ being the population difference (32). This shows that for all
$s\neq 0$, the squeezing factor $Q_z<1$, which implies that $S_z$ is 
squeezed with respect to $S_\pm$. The latter, in physical parlance, means 
the squeezing of atomic states.

\subsection{Multiparticle entanglement}

To study the possible entanglement for a multiparticle system, one has to 
consider the reduced density matrices. A $p$-particle reduced density 
matrix is given by 
\be
\label{52}
\rho_p(\br_1\ldots\br_p,\br_1'\ldots\br_p',t) \equiv {\rm Tr}\;
\psi(\br_1)\ldots\psi(\br_p)\hat\rho(t) \psi^\dgr(\br_p')\ldots
\psi^\dgr(\br_1')\; ,
\ee
where $\br_i$ are Cartesian coordinates, $t$ is time, $\psi(\br)$ and 
$\psi^\dgr(\br)$ are field operators, $\hat\rho(t)$ is a statistical operator,
and the trace is taken over the Fock space. As a basis for calculating the 
trace, we may accept coherent states [48] or Fock-Hartree states [59,60]. At 
low temperature, when practically all $N$ atoms are Bose-condensed, the trace 
(52) is concentrated on the coherent states normalized to the number of 
particles $N$. These coherent states have the form of the column
\be
\label{53}
|\eta_{n\al}>\; = \left [ \frac{e^{-N/2}}{\sqrt{k!}}\; \prod_{k=1}^k
\eta_{n\al}(\br_i)\right ] \; ,
\ee
in which
\be
\label{54}
\eta_{n\al}(\br) =\sqrt{N}\; \vp_n(\br) \; e^{i\al} \qquad 
(0\leq \al\leq 2\pi) \; ,
\ee
$\vp_n(\br)$ being the solution to the eigenproblem (1). The phase $\al$ 
here is random, because of which the vector (53) may be called the 
random-phase coherent state or mixed coherent state [48,61]. These states 
are asymptotically orthogonal, 
$$
<\eta_{m\al}|\eta_{n\bt}>\; \simeq \dlt_{mn}\; \dlt_{\al\bt} \qquad
(N\gg 1)\; .
$$
The closed linear envelope
\be
\label{55}
{\cal F}_0 \equiv \overline{\cal L} \{|\eta_{n\al}>\}
\ee
forms a subspace of the Fock space. On this subspace, one has the asymptotic, 
as $N\ra\infty$, resolution of unity
\be
\label{56}
\sum_n \int |\eta_{n\al}><\eta_{n\al}|\; \frac{d\al}{2\pi} \simeq \hat 1 \; ,
\ee
understood in the weak sense.

Let us also define a Hilbert space ${\cal H}$ as a closed linear envelope 
${\cal H}\equiv\overline{\cal L}\{\vp_n(\br)\}$ supplemented with a scalar 
product. And let us introduce a $p$-fold tensor product ${\cal H}^p\equiv
{\cal H}\otimes{\cal H}\otimes\ldots\otimes{\cal H}$. Then the functions
$$
\vp_n^p(\br_1,\ldots,\br_p) \equiv \prod_{i=1}^p \vp_n(\br_i)
$$
are approximately orthogonal on ${\cal H}^p$ in the following sense. The scalar
product
$$
\left ( \vp_m^p,\vp_n^p\right ) = \prod_{i=1}^p (\vp_m,\vp_n)
$$
has the property
$$
0\leq \left |\left ( \vp_m^p,\vp_n^p\right )\right | \leq 1 \; ,
$$
being equal to unity if and only if $m=n$. But for $m\neq n$, one has
$$
\lim_{p\ra\infty} \left |\left ( \vp_m^p,\vp_n^p\right )\right | = 0 \qquad
(m\neq n) \; .
$$

Assuming that the trace in Eq. (52) is concentrated on the condensate subspace (55),
we find
\be
\label{57}
\rho_p(\br_1\ldots\br_p,\br_1'\ldots\br_p',t) \cong \sum_n D_n(t)
\prod_{i=1}^p \vp_n(\br_i)\; \vp_n^*(\br_i') \; ,
\ee
where the coefficients are
\be
\label{58}
D_n(t) \equiv \; <\eta_{n\al}\; |\hat\rho(t)|\; \eta_{n\al}> \; .
\ee
The latter can be defined from the normalization conditions
$$
\int \rho_p(\br_1\ldots\br_p,\br_1\ldots\br_p,t) \; d\br_1\ldots
d\br_p = \frac{N!}{(N-p)!} \; , \qquad \sum_n|c_n(t)|^2 = 1 \; ,
$$
which suggests
\be
\label{59}
D_n(t) = \frac{N!}{(N-p)!}\; |c_n(t)|^2 \; .
\ee
For the resonant Bose condensate, with $N\gg 1$, one gets
\be
\label{60}
D_1(t) \simeq \frac{1-s}{2}\; N^p \; , \qquad 
D_2(t) \simeq \frac{1+s}{2}\; N^p \; .
\ee

The density matrix (57) cannot be presented as a product of single-particle 
density matrices, which implies that the state of the system is entangled 
[62,63]. This happens for all $|s|\neq 1$. Maximal entanglement occurs at 
$s=0$, when $D_1(t)=D_2(t) \simeq N^p/2$.

\section{Conclusion}

The method of resonant formation of nonlinear coherent modes makes it 
possible to create a novel type of systems, the resonant Bose condensate.
The latter shares many analogies with a resonant atom. But, being  a 
collective system, the resonant Bose condensate also displays the features 
essentially distinguishing it from a single resonant atom. First of all, the 
spectrum of the nonlinear coherent modes is described by the stationary 
Gross-Pitaevskii equation, which is a nonlinear Schr\"odinger equation. These 
modes, though formally looking as single-particle functions, correspond to 
collective states of a coherent multiatomic system. It would be admissible 
to say that such coherent modes are {\it single-quasiparticle} wave functions,
but not single-particle ones. Here, one should imply a quasiparticle in the 
sense of Landay, that is, an effective dressed single object inside a 
multiparticle ensemble.

The evolution equations for the resonant Bose condensate can be reduced to 
the system of two equations for the complex amplitudes of two fractional 
populations. This reduction to an effective two-mode problem is analogous
to that happening for a resonant atom. As in the case of any two-mode, 
two-level, or two-component system, the problem allows for the usage 
of quasispin representation. Employing this representation, one could 
talk about quasispin dynamics, qiasispin waves [64], having some formal 
properties similar to the dynamics of real spins [65], and so on. To study 
phase portraits, it is also convenient to work in the representation 
employing two real quantities: population difference and phase difference.

The collective nonlinear nature of the resonant Bose condensate yields the
existence of several novel effects making this system rather different 
from a single resonant atom. The most interesting of these effects, we have
investigated, are: {\it mode locking}, {\it critical dynamics}, {\it 
interference effects}, {\it atomic squeezing}, and {\it multiparticle 
entanglement}. These effects can find a broad range of various applications.

\vskip 5mm

{\bf Acknowledgement}

\vskip 2mm

One of the authors (V.I.Y.) appreciates useful discussions with K.P. Marzlin.

The main part of the work has been accomplished in the Research Center for
Optics and Photonics at the University of S\~ao Paulo, S\~ao Carlos. Financial
support from the S\~ao Paulo State research Foundation is acknowledged.

\newpage

\begin{center}

{\large{\bf Figure captions}}

\end{center}

\vskip 1cm

{\bf Fig. 1}. Local residual (12) as a function of the dimensionless radial 
variable $r$ for the self-similar approximant (solid line), optimized Gaussian 
approximation (dotted line), and Thomas-Fermi approximation (dashed line) for 
different values of the coupling parameter: (a) g=25; (b) g=250. The residual 
for the Thomas-Fermi approximation diverges at the edge of the atomic cloud.

\vskip 1cm

{\bf Fig. 2}. Atomic density (13) as a function of $r$ for the self-similar
approximant (solid line), optimized Gaussian approximation (dotted line), and 
Thomas-Fermi approximation (dashed line), at the coupling parameter $g=2500$.

\vskip 1cm

{\bf Fig. 3}. Percentage errors of the self-similar root approximants 
$e_1^*(g)$ (solid line), $e_2^*(g)$ (dashed line), and $e_3^*(g)$ (dotted 
line), as functions of the coupling parameter $g$, for vortex nonlinear modes 
with $n=j=0$ and different winding numbers $m$ and for different trap aspect 
ratio: (a) $\nu=1$, $m=1$; (b) $\nu=1$, $m=10$; (c) $\nu=10$, $m=0$; (d)
$\nu=10$, $m=1$; (e) $\nu=10$, $m=10$; (f) $\nu=100$, $m=1$; (g) $\nu=100$,
$m=10$.

\vskip 1cm

{\bf Fig.4}. Phase portrait on the plane of the variables $s(t)$ and $x(t)$
for the detuning $\ep=-0.1$ and the amplitude $b=0.51$ that is larger than 
the critical value $b_c=0.39821$ associated with the initial conditions $s_0=-1$,
$x_0=0$.

\end{document}